\documentclass[aps,prb,reprint,superscriptaddress]{revtex4-2}

\usepackage{graphicx}
\usepackage{amsmath}
\usepackage{comment}
\usepackage{braket}
\usepackage{xcolor}
\usepackage{hyperref}
\hypersetup{
    colorlinks=true,
    citecolor=teal,
    linkcolor=teal,
    filecolor=teal,      
    urlcolor=teal,
}

\usepackage[nameinlink,capitalise]{cleveref}

\usepackage[normalem]{ulem}

\newcommand{\be}{\begin{eqnarray}}
\newcommand{\ee}{\end{eqnarray}}

\begin{document}

\title{Spin and Charge Conductivity in the Square Lattice Fermi-Hubbard Model}

\author{Linh Pham }
\affiliation{Department of Physics and Astronomy, San Jos\'e State University, San Jos\'e, California 95192, USA}
\affiliation{Department of Physics and Astronomy, University of California, Davis, California 95616, USA}
\author{Ehsan Khatami}
\email[]{ehsan.khatami@sjsu.edu}
\affiliation{Department of Physics and Astronomy, San Jos\'e State University, San Jos\'e, California 95192, USA}

\date{\today}

\begin{abstract}
Dynamical properties are notoriously difficult to compute in numerical treatments of the Fermi-Hubbard model, especially in two spatial dimensions. However, they are essential in providing us with insight into some of the most important and less well-understood phases of the model, such as the pseudogap and strange metal phases at relatively high temperatures, or unconventional superconductivity at lower temperatures, away from the commensurate filling. Here, we use the numerical linked-cluster expansions to compute spin and charge optical conductivities of the model at different temperatures and strong interaction strengths via the exact real-time-dependent correlation functions of the current operators. We mitigate systematic errors associated with having a limited access to the long-time behavior of the correlators by introducing fits and allowing for non-zero Drude weights when appropriate. We compare our results to available data from optical lattice experiments and find that the Drude contributions can account for the theory-experiment gap in the DC spin conductivity of the model at half filling in the strong-coupling region. Our method helps paint a more complete picture of the conductivity in the two-dimensional Hubbard model and opens the door to studying dynamical properties of quantum lattice models in the thermodynamic limit.
\end{abstract}

\maketitle

\section{Introduction}

Experiments with cold fermionic atoms in optical lattices have shed a new light on the dynamical properties of itinerant fermion  models~\cite{Strohmaier2007,Schneider2012,p_brown_18,m_nichols_19,p_brown_19,Xu2019,Prichard2025,l_kendrick_25}, essential for our understanding of the electronic properties of quantum materials. Yet their numerical studies are met with special challenges. Unbiased methods used for treating system sizes beyond those accessible to exact diagonalization~\cite{Dagotto_92,t_tohyama_05,h_nakano_07} either do not have direct access to real-time data, e.g., in the case of quantum Monte Carlo (QMC), work best in the weakly-interacting region, e.g., in case of diagramatic Monte Carlo~\cite{y_eom_25}, or perform rather poorly in dimensions higher than one, e.g., in case of the density matrix renormalization group (DMRG), although there have been promising results with small clusters~\cite{Shinjo_21} and other advancements and applications with tensor network techniques in two dimensions (see for example Refs.~\cite{s_paeckel_19,b_ponsioen_20,c_hubig_20,c_yang_19,b_ponsioen_22}). Dynamical mean field theory~\cite{a_georges_96} based approaches, on the other hand, have limited access to non-local spatial correlations~\cite{m_ferrero_09,e_gull_10,f_simkovic_24}.

In QMC, correlations obtained in the Matsubara frequency space have to be analytically continued to the real-frequency domain ~\cite{Scalapino_92,d_scalapino_93}, a problem that is ill posed and can lead to unreliable results. In addition, these calculations often suffer from the infamous fermion ``sign problem"~\cite{e_loh_90,p_henelius_00,v_iglovikov_15} in the most physically interesting parameter regions. Despite that, recent tour de force calculations using the determinant QMC method have led to remarkable results for the spectra~\cite{e_huang_17,p_brown_19}, conductivity~\cite{e_huang_18}, and other dynamical properties~\cite{w_wang_23} of Fermi-Hubbard models.

Here, we employ the numerical linked-cluster expansion (NLCE) method~\cite{M_rigol_06,b_tang_13b} and its extension for computing dynamical properties of equilibrium systems (we call it {\it dynamical NLCE} or dNLCE for short), introduced in Ref.~\cite{m_nichols_19} (see also Ref.~\cite{Richter_19}), to study finite-temperature spin and charge optical conductivities of the two-dimensional (2D) square lattice Fermi-Hubbard model (FHM) in the thermodynamic limit. We obtain the out-of-time correlation functions of the corresponding current operators in the real time domain and use Kubo's formula to obtain conductivities directly in the real frequency axis within the linear response theory. Through a combination of fits and sum rules, we are able to identify and reduce potential systematic errors associated with having access to limited times for the correlators.

Our results point to nonzero Drude weights for spin, and in the Fermi liquid region, charge conduction at intermediate temperatures. We find reasonable agreement between the charge conduction in the direct current (DC) limit and results from optical lattice experiments, including the linear-in-temperature functionality of the charge resistivity in the intermediate doping region, on par with other numerical methods. Our technique lends itself well to studies in the strong-coupling region of the model, where the strength of Coulomb interactions is much larger than the nearest-neighbor tunneling amplitude. Hence, we revisit the question of spin conduction in that limit and are able to close the theory-experiment gap observed in previous calculations for the DC spin conductivity at half filling. We discuss the essential role a nonzero Drude weight plays in that. 

The rest of the paper is organized as follows. We present the Fermi-Hubbard Hamiltonian in Sec.~\ref{sec:model}. In Sec.~\ref{sec:nlce}, we provide a brief introduction to the NLCE algorithm and discuss the dNLCE process for obtaining optical conductivities. Here, we  also benchmark time-dependent correlation functions against DMRG results in one dimension (1D) and motivate the use of fitting functions. Sec.~\ref{sec:results} is dedicated to presenting our main results, including comparisons to experimental data. Finally, in Sec.~\ref{sec:discuss}, we summarize our findings and discuss potential future studies of dynamical properties of quantum lattice models using the dNLCE.

\section{The Model}
\label{sec:model}

We are interested in investigating the FHM, whose Hamiltonian is
\begin{equation}\label{eq:Hubbard_N1}
H = -t \sum_{\langle i,j \rangle, \sigma} \left( c_{i \sigma}^\dagger c_{j \sigma}^{\phantom{\dagger}} 
+ \mathrm{h.c.} \right) + U \sum_{i} n_{i \uparrow} n_{i \downarrow}  - \mu \sum_{i,\sigma} n_{i \sigma},
\end{equation} 
where $c_{i \sigma}^\dagger$ ($c_{i \sigma}^{\phantom{\dagger}} $) is the creation (annihilation) operator for a fermion with spin flavor $\sigma = \uparrow,\downarrow$ on site $i = 1,2,\dots$ of the lattice, $n_{i \sigma} = c_{i \sigma}^\dagger c_{i \sigma}^{\phantom{\dagger}}$ is the number operator for flavor $\sigma$, $t$ is the nearest-neighbor hopping amplitude and $U$ ($>0$ repulsive) is the interaction strength. We work in the grand canonical ensemble and use the chemical potential $\mu$ to adjust the fermion density. We work in units where $\hbar=k_B = 1$, where $\hbar$ is Planck's constant divided by $2\pi$ and $k_B$ is the Boltzmann constant, and set $t=1$ as the unit of energy throughout the paper.

\section{The Numerical Linked-Cluster Expansions}
\label{sec:nlce}

We use the NLCE~\cite{M_rigol_06,b_tang_13b}, and its application to dynamical (time dependent) properties, dNLCE, to calculate real-time charge and spin current-current correlation functions of the FHM (see Refs.~\cite{m_nichols_19,Richter_19} for similar previous applications to the FHM and the XXZ model). We then Fourier transform the latter to obtain optical conductivities directly in the real frequency domain.

NLCE is a cluster expansion in the {\it thermodynamic limit} for extensive properties of a lattice model in which the terms of the expansion are contributions from finite clusters on the lattice. These contributions are calculated recursively using the inclusion-exclusion principle, starting from a single site and by growing the clusters on the lattice up to a certain size. All possible cluster shapes and topologies up to that size are considered in the series. Here, we rely on the exact knowledge of the property on each cluster from exact diagonalization. Because of the translational symmetry, only distinct clusters not related by translation need to be considered. One can show that the contribution from disconnected clusters is exactly zero. So, only the connected (or {\it linked}) clusters are considered. Moreover, we use isomorphism and other symmetry considerations to reduce the number of clusters for which the Hamiltonian is diagonalized. We refer the interested reader to Ref.~\cite{b_tang_13b} for the details of the algorithm.

We use a single site as the building block to grow the clusters in the series, i.e., the {\it site expansion}, in which the {\it order} refers to the maximum number of sites for which the contribution of all possible clusters have been considered. Most NLCE studies involving the 2D FHM so far have used up to 9 orders. Here, due to the fact that there are four fermion operators in the current-current correlation functions in which we are interested (see below) and the addition of the time dimension, we have been able to obtain results for up to order 8. In 1D, a single cluster contributes to each order and we have been able to go up to order 10 at half filling. However, since the properties are calculated using exact diagonalization at the level of individual clusters, we are able to produce all properties of interest on a temperature, and a dense chemical potential, grid in a single run (for a given $U/t$). We use numerical resummations techniques, specifically the Wynn algorithm~\cite{Wynn,b_tang_13b} with 3 cycles of improvement, to extend the convergence to lower temperatures and longer times than possible with the original series.

Averaging over the two orientations of each distinct cluster in the series that are related by a $\pi/2$ rotation, we calculate the uniform (${\bf q}=0$) time-dependent charge ($c$) and spin ($s$) current-current correlation functions along the $x$ direction, $\braket{j_{c/s}(\tau)j_{c/s}(0)}$, where $\tau$ represents {\it real time}, $j_{c/s}(\tau)=e^{i\tau H}j_{c/s}e^{-i\tau H}$, and the current operators are given by
\be
j_{c} &=& it \sum_{l,\sigma} (c^\dagger_{l+\hat{x}\sigma}c_{l \sigma} - c^\dagger_{l\sigma}c_{l+\hat{x} \sigma})\label{eq:j1}\\
j_{s} &=& it \sum_{l,\sigma} \sigma (c^\dagger_{l+\hat{x}\sigma}c_{l \sigma} - c^\dagger_{l\sigma}c_{l+\hat{x} \sigma}).
\ee

We write the optical conductivities as~\cite{mahan,d_scalapino_93,c_karrasch_14}
\be
\sigma_{c/s}(\omega) = 2\pi \mathcal{D}_{c/s} \delta(\omega)+ \sigma_{c/s}^{(reg)}(\omega),\nonumber
\ee
where $\mathcal{D}$ is the Drude weight (not to be confused with the diffusion coefficient) so that the real part of the ``regular" part of the conductivity can be expressed as~\cite{mahan,c_karrasch_14} 
\be
\textrm{Re }\sigma_{c/s}^{(reg)}(\omega) &=& \frac{-2}{\omega}  \label{eq:ImOnly}\\
&\times&\textrm{ Im} \int_0^\infty d\tau e^{i\omega \tau} \textrm{Im} \left<j_{c/s}(\tau)j_{c/s}(0)\right>_{(reg)}, \nonumber
\ee
which uses only the imaginary part of the correlation functions and where in $\left<j_{c/s}(\tau)j_{c/s}(0)\right>_{(reg)}$, we have subtracted off a potential Drude weight (see below). Using the Kubo-Martin-Schwinger condition, one can show that 
\be
\textrm{Re } \sigma_{c/s}^{(reg)}(\omega) &=& \frac{(1-e^{-\beta \omega})}{\omega}  \label{eq:ReandIm} \\
&\times& \textrm{Re}\int_0^\infty d\tau e^{i\omega \tau}\left< j_{c/s}(\tau)j_{c/s}(0)
\right>_{(reg)},\nonumber
\ee
which uses both the real and imaginary parts, and is another equation we use to estimate Re $\sigma_{reg}(\omega)$, as suggested in Ref.~\cite{c_karrasch_14}. $\beta=1/T$ is the inverse temperature. 

Combining Eqs.~(\ref{eq:ImOnly}) and (\ref{eq:ReandIm}), we can derive a third equation for Re $\sigma_{c/s}^{(reg)}(\omega)$,~\cite{Zwierlein}
\be
\textrm{Re }\sigma_{c/s}^{(reg)}(\omega) &=& \frac{2 \tanh\left(\frac{\beta \omega}{2} \right)}{\omega} \label{eq:ReOnly} \\
&\times& \textrm{ Re} \int_0^\infty d\tau e^{i\omega \tau} \textrm{Re} \left<j_{c/s}(\tau)j_{c/s}(0)\right>_{(reg)},\nonumber 
\ee 
which uses only the real part of the correlation function. Note that using Eq. (\ref{eq:ReOnly}), the Drude weight can be obtained from the long-time limit of the correlators:
\be
\mathcal{D}_{c/s}&=&\frac{\beta}{2} \lim_{\tau\to\infty}\ \textrm{Re}\left<j_{c/s}(\tau)j_{c/s}(0)\right>,
\ee
hence, $\left<j_{c/s}(\tau)j_{c/s}(0)\right>_{(reg)}=\left<j_{c/s}(\tau)j_{c/s}(0)\right>-2\mathcal{D}/\beta$.

We then take the $\omega\to 0$ limit of $\textrm{Re }\sigma_{c/s}^{(reg)}(\omega)$, and add any Drude weight using a Lorentzian for broadening the delta function, as the DC conductivity.

Given the exact time-dependent current-current correlations, Eqs.~(\ref{eq:ImOnly})-(\ref{eq:ReOnly}) are expected to all give the same results for $\textrm{Re }\sigma_{c/s}^{(reg)}(\omega)$. In practice, however, the convergence of the dNLCE does not extend to long enough times, and any time cut-off will potentially result in disagreements between conductivities calculated via the three different equations. To mitigate these errors, we first fit the current-current correlations to a function and use the fitted functions to perform Fourier transforms. 

\begin{figure}[t]
\includegraphics[width=\linewidth]{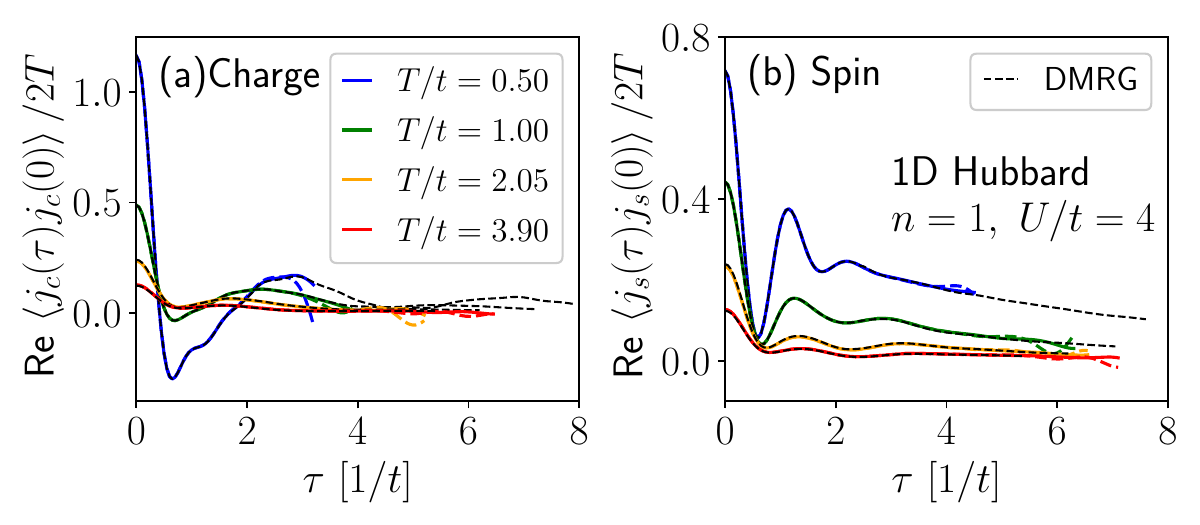}
\caption{Comparison of dNLCE current correlators of the FHM at different temperatures with those from DMRG in 1D. (a) The real part of the charge current-current correlation functions vs real time for the 1D FHM at half filling with $U/t=4$. Black dashed lines are the DMRG results for a $100$-site chain from Ref.~\cite{c_karrasch_14}. Solid and dashed color lines represent dNLCE results at orders 10 and 9, respectively (no resummations used). Their deviations from each other marks the time at which the series loses convergence. (b) Same as (a) except for the real part of the spin current-current correlation functions vs real time.}
\label{fig:dmrgcompare1} 
\end{figure}

We observe that the general form of the spin or charge current-current correlation functions can be described by damped oscillations with at least one frequency that is roughly set by the interaction strength $U/t$ (see Figs.~\ref{fig:dmrgcompare1}-\ref{fig:fit2dspincharge}). So, we consider the fitting function
\be
y(\tau) = \frac{2\mathcal{D}_{c/s}}{\beta} + Ae^{-B\tau}\sum_{n=1}^3 C_n \cos(E_n \tau - F_n)
\label{eq:expcos}
\ee
for the real part of the correlators, and the same function without the constant Drude term for the imaginary part. Here, $\mathcal{D}_{c/s}$, $A$, $B$, and $C_n$ are fitting parameters. Previously, an effective square {\it window function} was used in the Fourier transform, i.e., by setting the upper limit of integrals to a cut-off time~\cite{m_nichols_19}.

There can be several sets of fitting parameters that lead to near perfect matches of $y(\tau)$ to the correlation functions in the limited available time windows. So, as an additional condition for best fits, we pick those that also lead to minimum differences between conductivities calculated using Eqs.~(\ref{eq:ImOnly})-(\ref{eq:ReOnly}) across all frequencies. We take any remaining differences as a measure of systematic uncertainty in our conductivities. As we will see below, the fits suggest that there is a nonzero and temperature-dependent $\mathcal{D}_s$ in almost all cases, and a nonzero $\mathcal{D}_c$ at low enough temperatures and for dopings sufficiently away from half filling.

\section{Results}
\label{sec:results}

\subsection{Benchmarks in one dimension}

\begin{figure}[t]
\includegraphics[width=\linewidth]{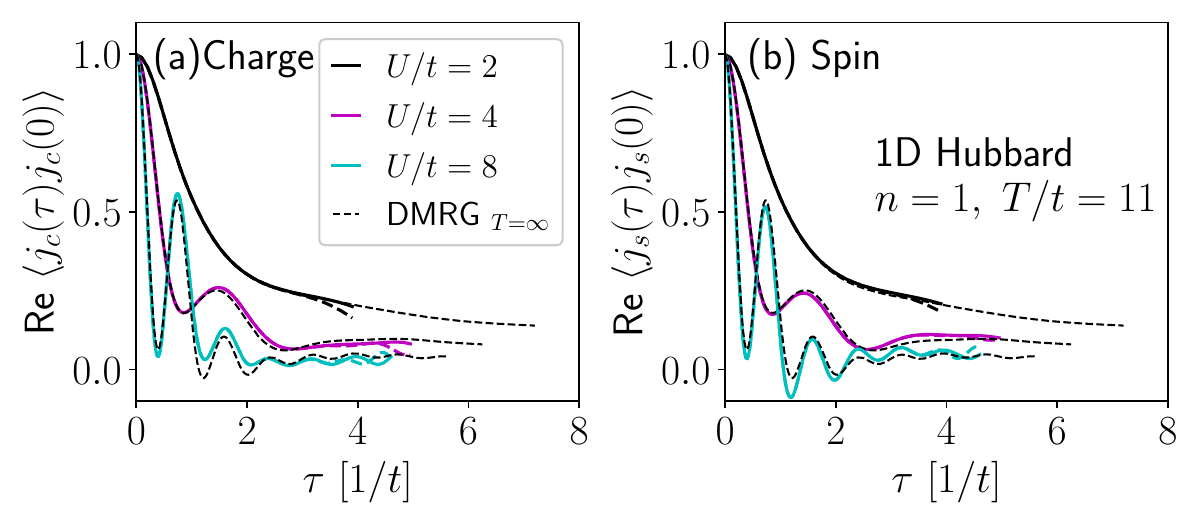}
\caption{Comparison of dNLCE current correlators of the half-filled FHM for different $U$ values with those from DMRG in 1D. Line types are the same as Fig.~\ref{fig:dmrgcompare1}, except that the DMRG results are at $T=\infty$ and dNLCE results are at a high temperature of $T/t=11$. The DMRG data are from Ref.~\cite{c_karrasch_14}.}
\label{fig:dmrgcompare2} 
\end{figure}

We begin by benchmarking our time-dependent current correlators against the DMRG results in 1D. We find that at times before the dNLCE convergence is lost, there is an excellent agreement between the dNLCE and the DMRG results at all temperatures. In Fig.~\ref{fig:dmrgcompare1}, we show the real part of the charge and spin correlation functions at half filling as a function of real time at various temperatures for $U/t=4$. The thin dashed lines are from time-dependent DMRG calculations of Ref.~\cite{c_karrasch_14} for a system with $100$ sites, which we take to be good estimates for the thermodynamic limit. Thick dashed and solid lines are the last two orders of the dNLCE results. The deviation of the latter two from each other indicates the loss of convergence in the dNLCE, which we observe to moves to earlier times as the temperatures is lowered. We find an excellent agreement with DMRG results within the convergence region. We find that numerical resummations do not lead to any improvements in the convergence of the series in the 1D case. Therefore, we have shown only the results for the bare series in this plot.

\begin{figure}[t]
    \includegraphics[width=\linewidth]{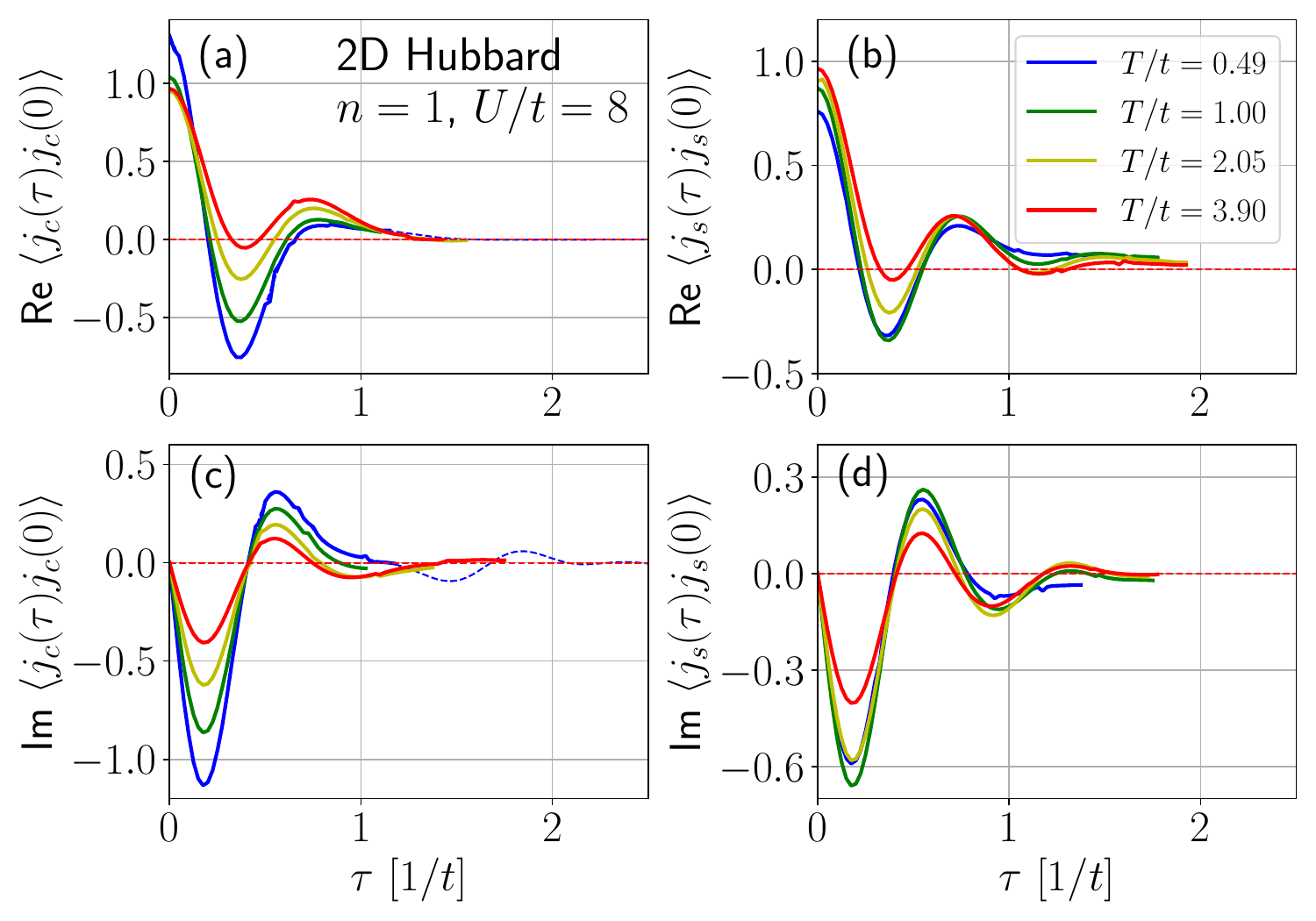}
    \caption{dNLCE results for the (a)-(b) real and (c)-(d) imaginary parts of the charge (left) and spin (right) current correlators of the half-filled 2D FHM vs real time at different temperatures for $U/t=8$. Solid lines show results after the Wynn resummation and dashed lines are fits based on Eq.~\ref{eq:expcos}. The dNLCE data are shown up to $\tau \sim 1.5/t-2.0/t$, roughly within their time interval of convergence. The same time interval is used in the fits. The resummations can result in spurious features in some regions, which show up as small dents, mostly at low temperatures.}
    \label{fig:fit2dspincharge}
\end{figure}

We also find a very good agreement between DMRG results for the same correlators at infinite temperature and the dNLCE calculations at the highest temperature in our grid ($T/t\sim 11$). In Fig.~\ref{fig:dmrgcompare2}, we show this for three different values of $U$ at half filling. Here, we have normalized the correlators to $1$ at $\tau=0$ for a better comparison across multiple $U$'s. There is a great overall agreement between the two methods. However, we find small quantitative disagreements at longer times that grow with increasing the interaction strength from $U/t=2$ to $U/t=8$. We attribute this to the finite temperature of the dNLCE results, which is comparable to the largest $U/t$.

\subsection{Current correlations in 2D}

In Fig.~\ref{fig:fit2dspincharge}, we show real and imaginary parts of the spin and charge current correlators for the half-filled FHM in the 2D square lattice as a function of real time for the intermediate interaction value of $U/t=8$. Fits to the data are also shown as thin dashed lines. We observe that the oscillations in time are quickly damped in all cases, especially for the charge current correlators. This is observed for all $U/t$ and densities we have studied in 2D. We cannot resolve any nonzero but small Drude weight that might exist in the charge channel for this $U/t$ at half filling. However, the real part of the spin correlators hint at the presence of a Drude weight that grows by decreasing the temperature, something we capture in our fits [see nonzero asymptotic values in Fig.~\ref{fig:fit2dspincharge}(b)].

\begin{figure}[t]
    \includegraphics[width=\linewidth]{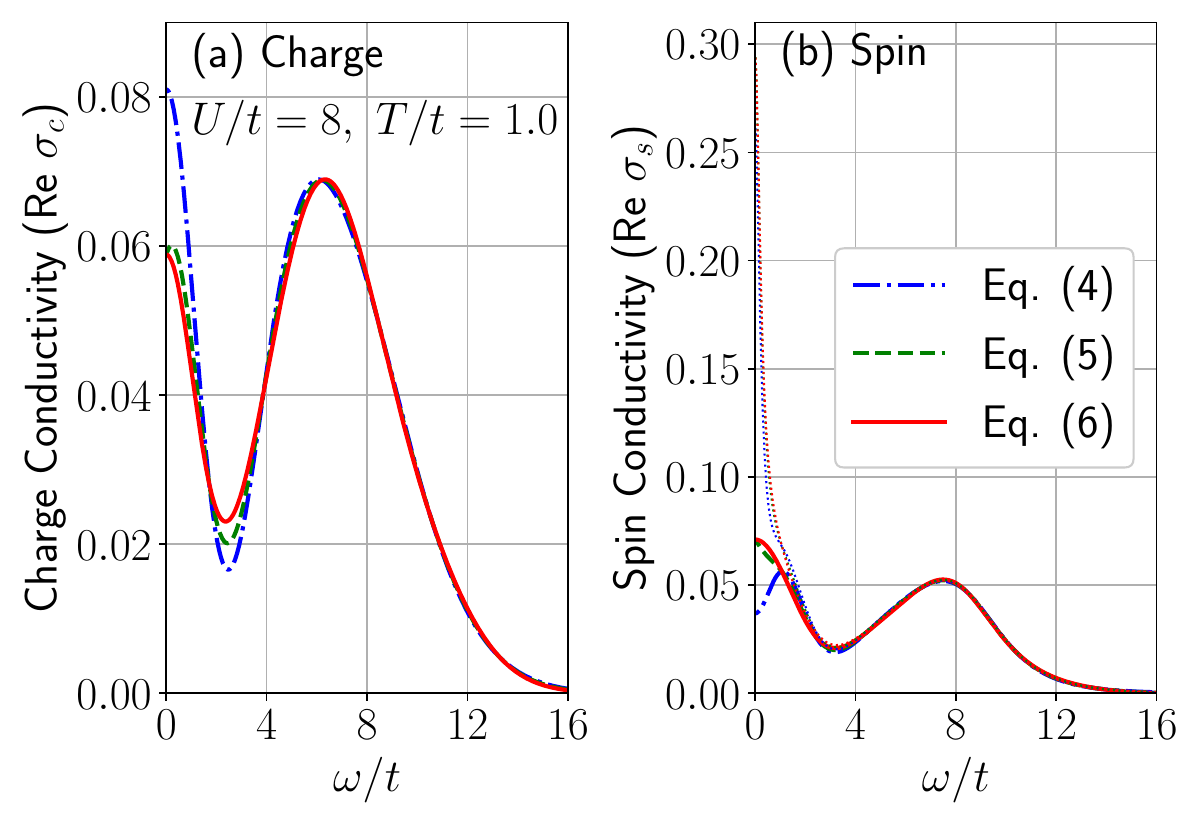}
    \caption{(a) Charge and (b) spin optical conductivities of the 2D FHM at half filling for $U/t=8$ and $T/t=1.0$, obtained from Fourier transforms of the corresponding current operators according to Eqs. (\ref{eq:ImOnly}) - (\ref{eq:ReOnly}). Thick lines in (b) are the regular parts of the spin conductivity (Re $\sigma_{s{\textrm (reg)}}$) and thin dotted lines are with the estimate for the Drude weight added (see text).}
    \label{fig:ACCond}
\end{figure}

Fourier transforming the anti-commutator of the correlation functions, we obtain three estimates for the real part of the corresponding charge or spin conductivities based on Eqs.~(\ref{eq:ImOnly})-(\ref{eq:ReOnly}) discussed above. Figure~\ref{fig:ACCond} shows the estimated curves at $T/t=1.0$ for the correlation functions shown in Fig.~\ref{fig:fit2dspincharge}. We find that those obtained from Eqs.~(\ref{eq:ReandIm}) or (\ref{eq:ReOnly}) more or less agree with each other and point to a larger peak in the DC limit of $\omega=0$ than the one obtained from Eq.~(\ref{eq:ImOnly}), which uses only the imaginary part of the correlators. The disagreements are a direct result of the lack of access to exact correlation functions beyond the convergence regions in the dNLCE and a measure of how well our extrapolations capture them. However, as suggested by our sum rules (see below), we will see that what is shown in Fig.~\ref{fig:fit2dspincharge} is in fact a representation of a worst case scenario.

In addition to the {\it regular} part of the spin conductivity, in Fig.~\ref{fig:ACCond}(b), we also show the spin conductivity when the Drude weight has been added (thin dotted lines) by replacing $\delta(\omega)$ with a Lorentzian of width $J=4t^2/U=0.5t$. The latter is the strength of the spin exchange interaction in the effective anti-ferromagnetic Heisenberg model that describes the physics of the half-filled FHM at strong interactions~\cite{t_paiva_01}. One expects to observe features that are relevant to this energy scale at time scales of $2\pi/J$, which are well beyond what we can achieve. The broadening of the delta function by $J$ is, therefore, an attempt to approximately account for some of the missing weight due to spin conduction processes that involve spin exchange.

\subsection{f-sum rule and other checks}

\begin{figure}[t]
    \includegraphics[width=\linewidth]{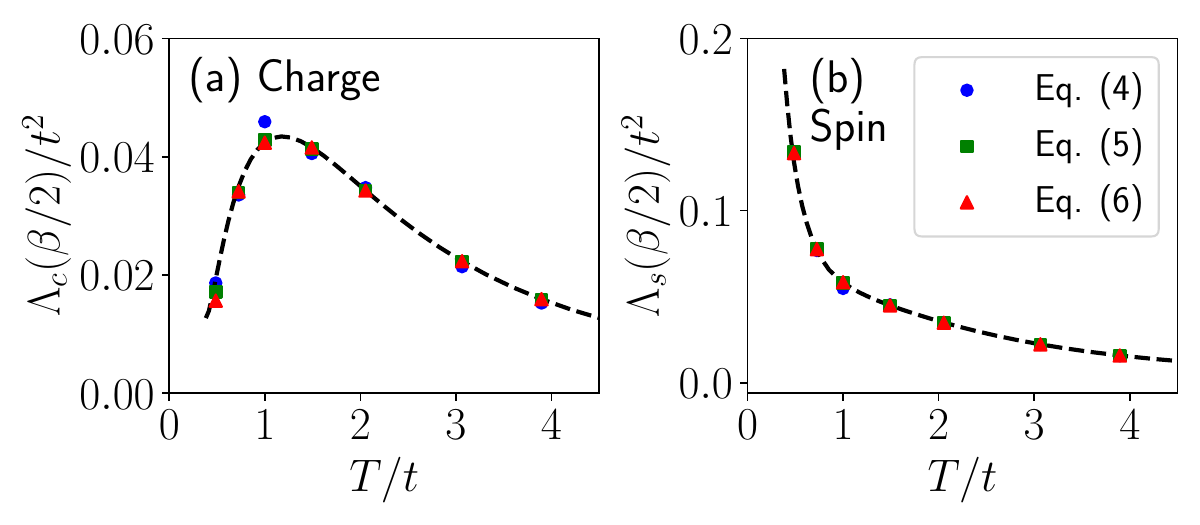}
    \caption{The imaginary time (a) charge and (b) spin current-current correlation functions of the FHM for $U/t=8$ at half filling at the midpoint of the inverse temperature range ($\beta/2$) as a function of temperature. The black dashed line is calculated directly in the NLCE and the markers are those obtained from the right hand side of Eq. (\ref{eq:lambdas}) with the corresponding optical conductivities from Eqs. (\ref{eq:ImOnly}) - (\ref{eq:ReOnly}) (see Fig.~\ref{fig:ACCond}). The agreements serve as a check on the validity of the optical conductivities.}
    \label{fig:dccheck}
\end{figure}

To independently verify which one of the estimates of the optical conductivity are more trustworthy, we use two integral equations (sum rules). The first one relates the optical conductivity to the current-current correlator in the imaginary time, or the retarded Green's function of the current operator in the imaginary time space, $\Lambda_{c/s}(\mathcal{T}) = - \left<T_{\mathcal{T}} j_{c/s}(\mathcal{T})j_{c/s}(0)\right>$, where $\mathcal{T}$ is the imaginary time and $T_{\mathcal{T}}$ is the imaginary time ordering operator~\cite{mahan,n_trivedi_96},
\be
\Lambda_{c/s}(\mathcal{T}) = \int_{-\infty}^{\infty} \frac{d\omega}{\pi} \frac{\omega e^{-\mathcal{T}\omega}}{1-e^{-\beta\omega}} \textrm{ Re }\sigma_{c/s}(\omega). \nonumber
\ee
At $\mathcal{T}=\beta/2$, the integral simplifies to
\be
\Lambda_{c/s}(\beta/2) &=& \int_{-\infty}^{\infty} \frac{d\omega}{\pi}\  \frac{\omega}{2\sinh (\beta\omega/2)}  \textrm{ Re }\sigma_{c/s}(\omega), \label{eq:lambdas} 
\ee
which at low enough temperatures, can be approximated to be $\frac{\pi}{\beta^2}\textrm{ Re }\sigma_{c/s}(0)$. Hence, $\frac{\beta^2}{\pi}\Lambda_{c/s}(\beta/2)$ is often used as a proxy for the DC conductivity in quantum Monte Carlo studies of the model as a way to avoid analytic continuation to the real frequency space~\cite{n_trivedi_96,p_denteneer_99}. However, it has been found to underestimate the actual charge or spin DC conductivities~\cite{e_huang_18,m_nichols_19}. It is worth noting that a similar approximation can be made for the spectral function~\cite{n_trivedi_95}.

We calculate the left hand side of Eq.~(\ref{eq:lambdas}) directly and exactly within the NLCE, and independently, calculate the right hand side using the three functions we obtain for $\textrm{Re } \sigma_{c/s}(\omega)$. In Fig.~\ref{fig:dccheck}, we show the results as a function of temperature for the same model parameters as used in Figs.~\ref{fig:fit2dspincharge} and \ref{fig:ACCond}. $\Lambda_{c/s}(\beta/2)$ are shown as dashed lines and the integrals are shown as symbols. We observe a generally very good agreement between the two. This agreement does not seem to depend much on which Fourier transform is used to obtain $\textrm{Re } \sigma_{c/s}(\omega)$. The only visible exceptions in this case are at $T/t=1.0$ and $T/t=0.49$, especially for charge in Fig.~\ref{fig:dccheck}(a). Specifically, we find that at $T/t=1.0$, the results obtained from Eqs.~(\ref{eq:ReandIm}) and (\ref{eq:ReOnly}) exhibit a better agreement with $\frac{\beta^2}{\pi}\Lambda_c(\beta/2)$, suggesting that the $\textrm{Re }\sigma_c(\omega)$ obtained from Eq.~(\ref{eq:ImOnly}) in Fig.~\ref{fig:ACCond} is less accurate than the other two estimates. This effect is also present, albeit not as pronounced, in Fig.~\ref{fig:dccheck}(b) for spin at the same temperature. On the other hand, we find that at $T/t=0.49$, the charge conductivities obtained from Eqs.~(\ref{eq:ReandIm}) and (\ref{eq:ReOnly}) perform worse for this check than that from Eq.~(\ref{eq:ImOnly}). However, this appears to be merely a coincident. Examining the optical charge conductivities at that temperature (not shown) reveals that there are significant differences between the results from the three integrals starting at frequencies larger than $U/t$ and that none may be trusted. Hints of that could be found in the suddent change of character of the best fit we find to the imaginary part of the corresponding current correlator [see the blue curves in Fig.~\ref{fig:fit2dspincharge}(c)].

\begin{figure}[t]
    \includegraphics[width=\linewidth]{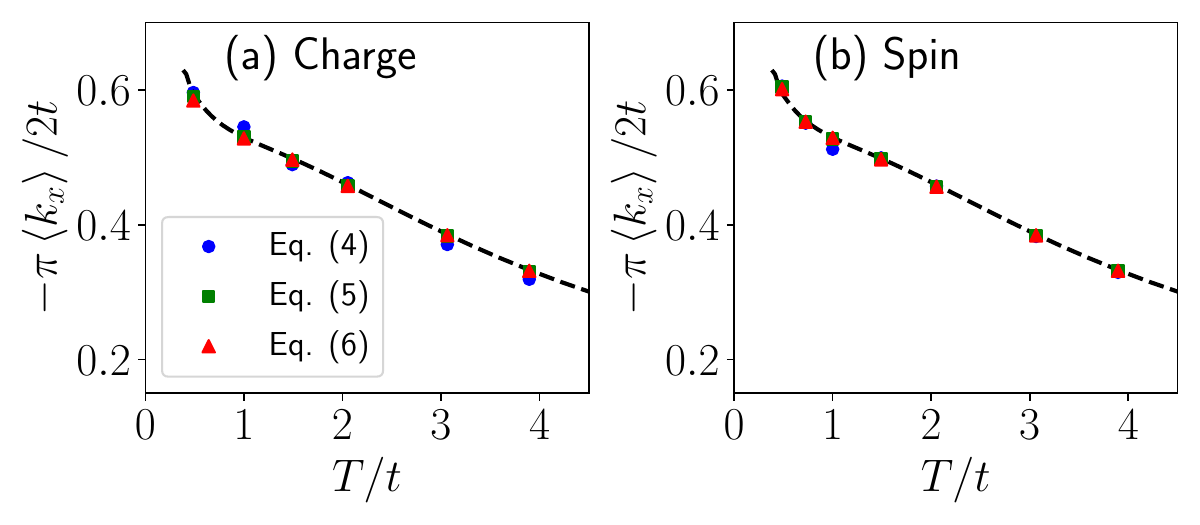}
    \caption{Check of the f-sum rule [Eq. (\ref{eq:fsumrule})] for (a) charge and (b) spin conductivities  of the FHM for $U/t=8$ at half filling as a function of temperature. The black dashed lines are directly calculated in the NLCE and the markers are calculated using the integral on the right hand side of Eq. (\ref{eq:fsumrule}) with corresponding optical conductivities from Eqs. (\ref{eq:ImOnly}) - (\ref{eq:ReOnly}) (see Fig.~\ref{fig:ACCond}). Note that the Drude weight for the spin conductivity is taken into account in the integral.}
    \label{fig:fsumrule}
\end{figure}

The second integral equation we use to assess which one of the estimates for our optical conductivity is likely the most accurate one is the so-called f-sum rule~\cite{r_fishman_02}, which relates the integral of the real part of the conductivity (for both spin and charge) to the kinetic energy along the direction of the current, namely,
\be
-\frac{\pi \langle k_x \rangle}{2} &=&\int_0^{\infty} \textrm{Re} \, \sigma_{c/s}(\omega) d\omega.
\label{eq:fsumrule}
\ee
Whereas the function of $\omega$ that multiplies Re $\sigma(\omega)$ in the integral of Eq.~(\ref{eq:lambdas}) gives more weight to the disputed small frequency region of Re $\sigma(\omega)$, the integral in Eq.~(\ref{eq:fsumrule}) does not contain such reweighting of the optical conductivity. Nevertheless, in Fig.~\ref{fig:fsumrule}, we investigate its validity by comparing the exact kinetic energy on the left hand side of the equation (dashed line) to the values obtained from the integral on the right hand side, based on the three equations for the conductivities, as a function of temperature. We find that, similar to what we observed in Fig.~\ref{fig:dccheck}, the Fourier transforms that use the real part of the correlation function [Eqs.~(\ref{eq:ReandIm}) and (\ref{eq:ReOnly})] lead to the best agreement across all temperatures. We find that this is also consistently the case across all the $U$ values and densities we have analyzed. Hence, we will work with the $\sigma_{c/s}(\omega)$ obtained from Eq.~(\ref{eq:ReOnly}) as the most reliable estimate for the optical conductivity in the remainder of the paper.

\begin{figure}[t]
    \includegraphics[width=\linewidth]{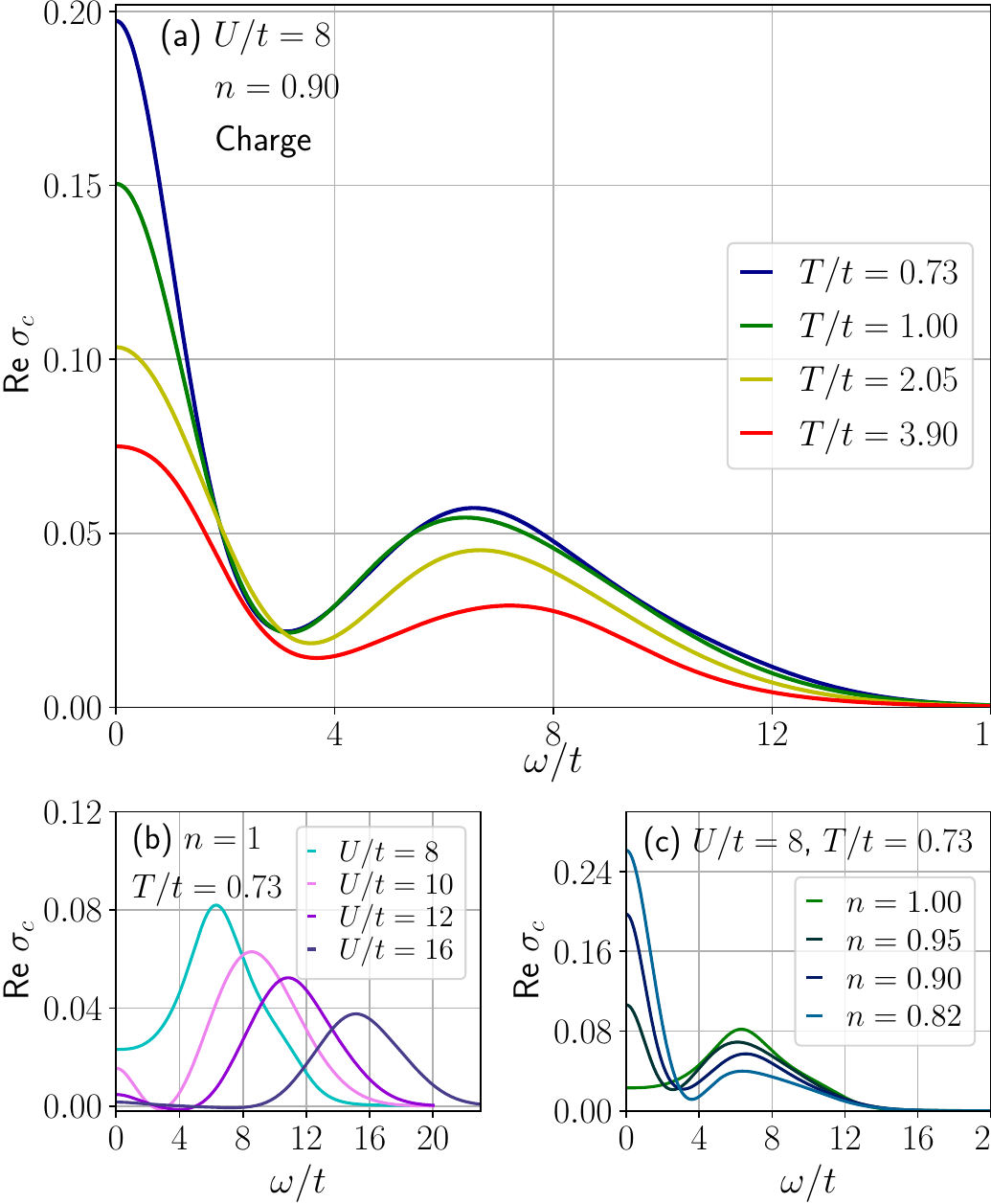}
    \caption{(a) Optical charge conductivity of the 2D FHM for $U/t=8$ at the density of $n=0.90$ and different temperatures, obtained from Eq.~\ref{eq:ReOnly}. The DC peak grows as the temperature decreases. Another pronounced peak develops at $\omega/t$ between $6$ and $8$. (b) Same optical conductivity as in (a), except at half filling and $T/t=0.73$, for four different $U/t$ values. The DC conductivity is suppressed and the high-frequency peak moves to larger frequencies, tracking $U/t$. (c) The evolution of the same optical conductivity as in (a), except at $T/t=0.73$ for several densities at, and away from, half filling. The DC conductivity grows significantly by increasing the doping.}
    \label{fig:ACCharge}
\end{figure}

\begin{figure}[t]
    \includegraphics[width=\linewidth]{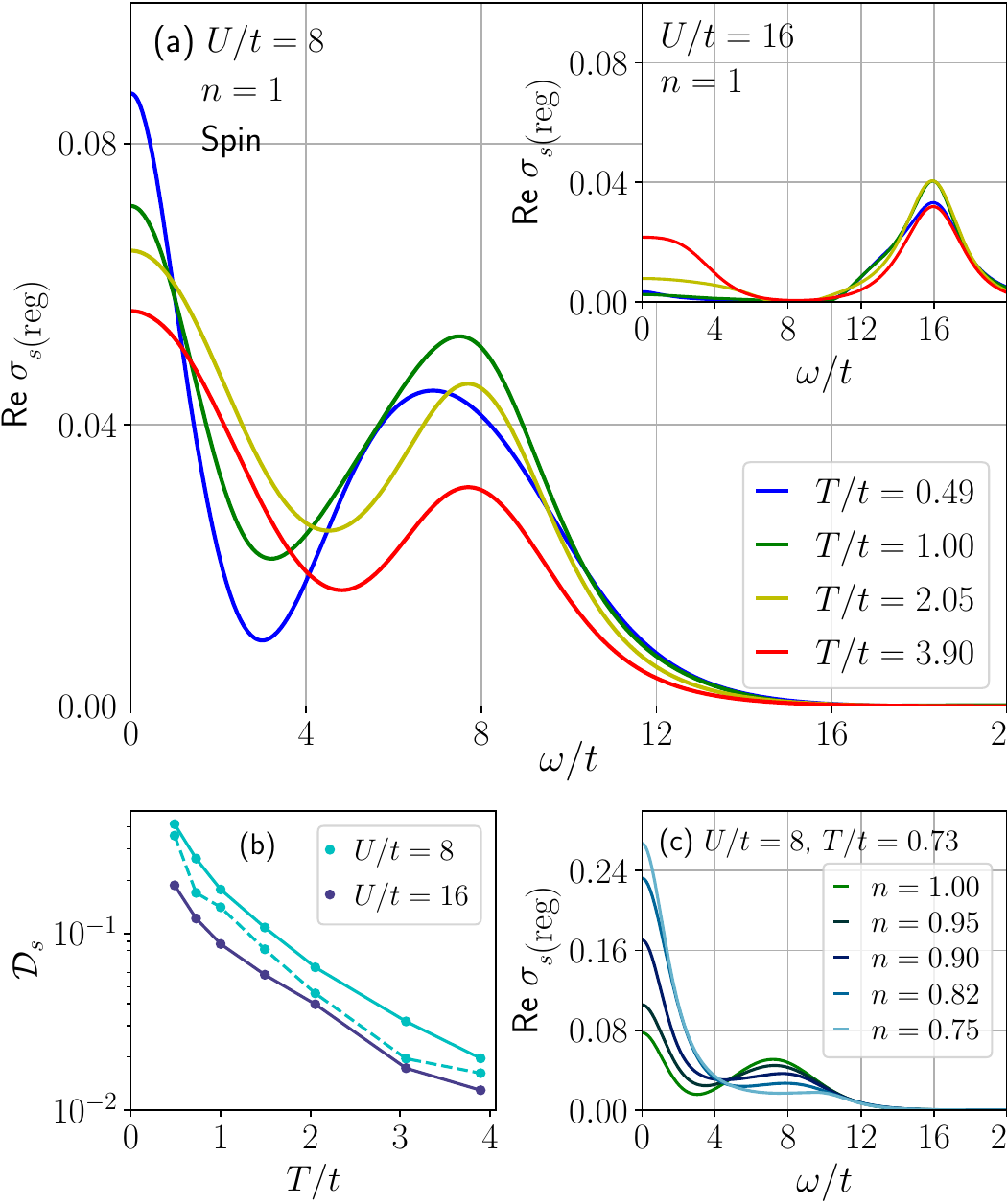}
    \caption{(a) The regular part of optical spin conductivity of the 2D FHM for $U/t=8$ at half filling and different temperatures, obtained from Eq.~\ref{eq:ReOnly}. The high-frequency peak is again around $U$, and the DC peak grows slightly as the temperature decreases. The inset shows the same quantity for $U/t=16$. The DC limit of the conductivities are largely suppressed in comparison to those for $U/t=8$. (b) The corresponding Drude weights as a function of temperature. (c) Same as in (a), except at $T/t=0.73$ for several densities at, and away from, half filling. Like for charge, the DC limit grows significantly by increasing the doping.}
    \label{fig:ACSpin}
\end{figure}

\subsection{Optical conductivities}

The charge optical conductivity shows a high-frequency peak around the characteristic frequency $\omega\sim U$ and a Drude-type peak that grows with decreasing temperature or $U/t$. In Fig.~\ref{fig:ACCharge}(a), we show this for $U/t=8$ at 10\% doping away from half filling ($n=0.90$) and at four different temperatures below and above the hopping amplitude. The high-frequency peak, which is just below $\omega=U$ at the high temperature of $T/t=3.9$, moves to slightly lower frequencies by lowering the temperature, but remains more or less unchanged for $T\lesssim 1.0$. At the same time, the Drude-type peak at $\omega=0$ grows by decreasing temperature, signaling a metallic behavior. Figure~\ref{fig:ACCharge}(b) shows the same quantity but at $n=1$ for $U/t=8-16$. Here, the decreasing DC conductivity by increasing $U/t$ reflects the widening of the charge gap, which is expected to fully develop at lower temperatures~\cite{n_bulut_94}, and an insulating behavior at half filling. As expected, the high-frequency peak tracks $U/t$. 

In stark contrast to what was observed at half filling, moving away from half filling at relatively low temperatures the system displays a Drude-like peak and turns increasingly metallic. This is clearly seen in Fig.~\ref{fig:ACCharge}(c), where the DC limit of the optical conductivity at $T/t=0.73$ is shown to rise rapidly and grow as the system is doped away from half filling. In fact, we find a clear indication of a nonzero Drude weight in the charge current-current correlation functions at $n<0.80$ (see below). Note that the width of the Drude-like feature does not change much upon doping. At $n=0.82$, much of the spectral weight around $\omega=U$ is transferred to the DC peak. This is expected since by increasing the doping beyond $20\%$ or so, one approaches the Fermi liquid region of the FHM.

Unlike for charge, the regular part of spin optical conductivity at half filling shows a strong signal in the DC limit, which grows with decreasing temperature [see Fig.~\ref{fig:ACSpin}(a)]. As $U$ increases beyond $U/t\sim 8$, the strength of spin exchange, which is the main driver of spin conduction at this filling, decreases as $J= 4t^2/U$. As a result, as shown in the inset of Fig.~\ref{fig:ACSpin}(a), we find that the strength of this signal is greatly reduced when $U/t=16$ in comparison to when $U/t=8$. However, we also find a significant enhancement of the conductivity in the DC limit upon doping away from the commensurate filling, as shown in Fig.~\ref{fig:ACSpin}(c), suggesting that both spin exchange and hopping are essential processes for spin conduction in the model, as was found to be the case for the corresponding low-energy $t-J$ model~\cite{j_bonca_95}. 

As mentioned above, our fits to the time-dependent current correlators yield a nonzero Drude weight for spin. We show the temperature dependence of this weight at half filling for $U/t=8$ and $16$ in Fig.~\ref{fig:ACSpin}(b). The dashed line in that figure shows $\mathcal{D}_s$ at 10\% doping for $U/t=8$. We find that $\mathcal{D}_s$ is strongly dependent on temperature, is larger for smaller $U/t$, and does not appear to have a dramatic dependence on doping, at least close enough to half filling.

\subsection{Charge and spin resistivity}

\begin{figure}[t]
    \includegraphics[width=\linewidth]{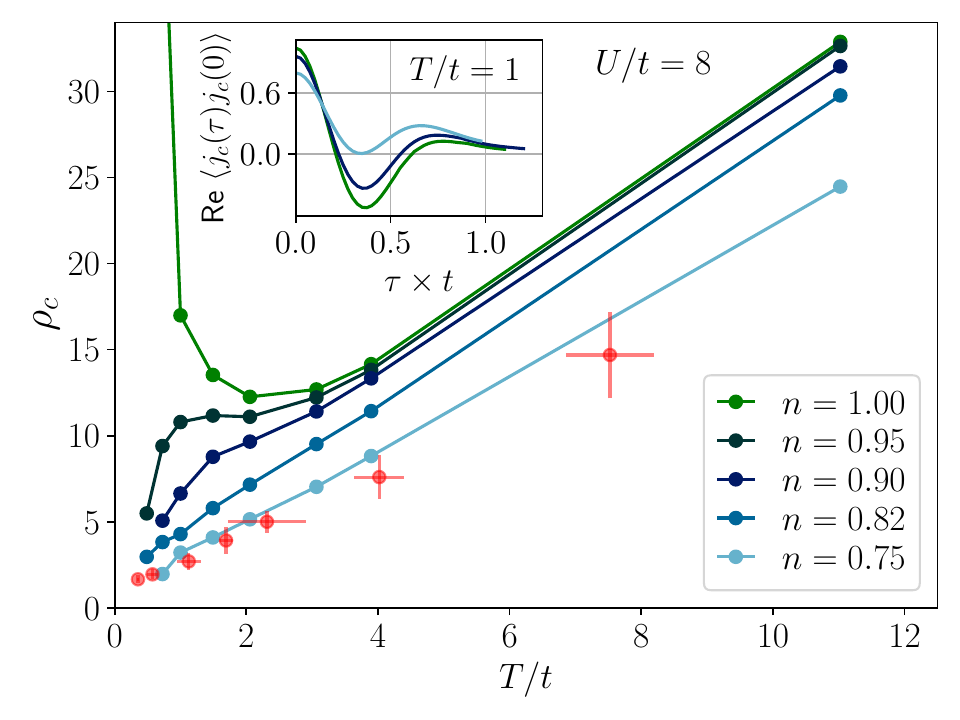}
    \caption{Charge resistivity of the 2D FHM [1/Re $\sigma_c(0)$] for $U/t=8$ vs temperature at several densities. A finite Drude weight is considered only in the Fermi liquid region with $n=0.75$ and at the lowest temperature when $n=0.95$. A clear linear-in-$T$ scaling is observed at the intermediate density of $n=0.82$. Red markers are experimental results for $U/t=7.5(8)$ and $n\sim 0.80$ from Ref.~\cite{p_brown_18}. The inset shows the real part of the time-dependent charge current-current correlation functions as a function of real time at $T/t=1$ and at densities $n=1.00$, $0.90$, and $0.75$.}
    \label{fig:ChargeRes}
\end{figure}

Turning our focus now exclusively to the DC limit, in Fig.~\ref{fig:ChargeRes}, we plot the DC charge resistivity $\rho_c=1/\sigma_c(\omega=0)$ for $U/t=8$ and densities at, and away from, half filling as a function of temperature. Regardless of the density, we find that at high temperatures ($T/t>2$), $\rho_c$ continues to increase linearly with increasing the temperature, appearing to violate the Mott-Ioffe-Regel limit~\cite{ioffe1960non,m_mott_72}. However, this ``bad metallic" behavior has been attributed to the temperature dependence of the charge susceptibility~\cite{j_kokalj_17}. 

At half filling, there is a sharp upturn below $T/t=2$, which is a clear signature of the Mott insulating behavior of the model at the commensurate filling. Doping the system even slightly away from that limit changes this behavior and turns the system into a metal. This is seen, for example, as the sudden downturn in $\rho_c$ at $T/t<1$ when $n=0.95$. At smaller densities, the decrease in resistivity is monotonic with temperature.

Specifically, for $n\sim 0.82$, we find that the almost perfect linear dependence of $\rho_c$ on the temperature extends to $T/t<2$, as expected in the ``strange metal" region~\cite{Hussey_2008} of the model, with resistivities at other densities showing a similar linear behavior at low temperatures, albeit with different slopes. This signature linear dependence in temperature has been previously observed in exact diagonalization study of small clusters~\cite{j_riera_94}, determinant quantum Monte Carlo simulations of the model with $U/t=6$ and a nonzero diagonal hopping amplitude of $t'/t=-0.25$~\cite{e_huang_18}, as well as in simulations involving cold fermionic atoms in optical lattices where $U/t=7.5(8)$ and $n\sim 0.80$~\cite{p_brown_18}. In fact, we plot the data from the latter experiment as red markers with error bars in  Fig.~\ref{fig:ChargeRes} for comparison. The deviation of our results from the experimental data can be attributed to the slight difference in the values of $U/t$, but also a potential systematic underestimation of the conductivity at low frequencies in dNLCE due to lack of access to exact long-time information in the real-time correlations.

While we cannot rule out a nonzero Drude weight for charge at any density away from half filling, at dopings above 20\%, we find that the real part of the charge current-current correlator vs time oscillates between values that remain above zero (see the inset of Fig.~\ref{fig:ChargeRes}), strongly suggesting that there exists a nonzero Drude weight, as expected, in the Fermi liquid phase of the model~\cite{Dagotto_92,g_uhrig_95}. Therefore, we allow for this in our fits of the current correlator at $n=0.75$. The DC resistivity shown in Fig.~\ref{fig:ChargeRes} at this density reflects a Drude peak that has been broadened via a Lorentzian using the hopping amplitude as the width. While $t$ is the most relevant energy scale when it comes to charge transport in the model, we note that this is a less-than-ideal choice since one would expect correlations at time scales $\mathcal{O}(1/t)$ to have already been captured by our time-dependent charge correlators. The width of the actual Drude weight being likely significantly smaller than $t$ suggests that our DC conductivities in the Fermi-liquid region are underestimations. We note that the persistence of the linear-in-temperature functionality of the resistivity in the Fermi-liquid region is consistent with previous numerical results for the model with a nonzero but small $t'$~\cite{d_bergeron_11,e_huang_18}.

\begin{figure}[t]
    \includegraphics[width=\linewidth]{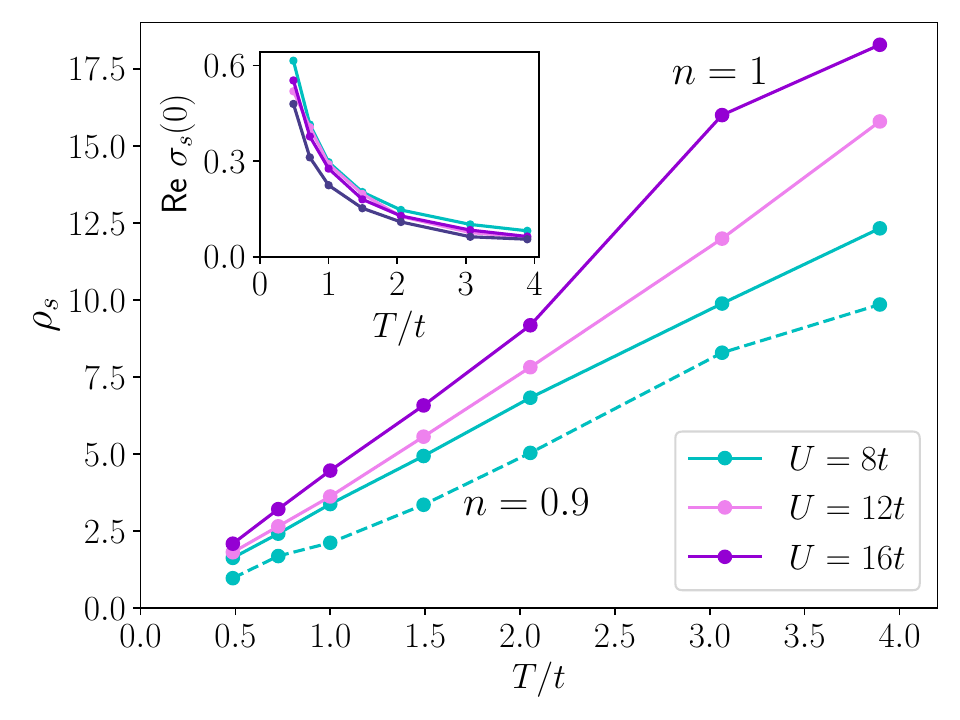}
    \caption{Spin resistivity of the 2D FHM [1/Re $\sigma_s(0)$] at half filling (solid lines) vs temperature for several $U/t$ values and at $n=0.90$ (dashed line) for $U/t=8$. The inset shows the DC conductivity [Re $\sigma_s(0)$] as a function of temperature for the same parameters.}
    \label{fig:spinresist}
\end{figure}

Turning to spin resistivity, $\rho_s=1/\textrm{Re}\ \sigma_s(0)$, we find that it decreases approximately linearly with lowering the temperature. In Fig.~\ref{fig:spinresist}, we show this quantity at half filling (solid lines) for three $U/t$ values of $8$, $12$, and $16$ in the strong-coupling region, and at the density of $n=0.90$ for $U/t=8$ (dashed line). As suggested by the corresponding optical conductivity in Fig.~\ref{fig:ACSpin}, increasing $U/t$ beyond $8$ in the strong-coupling region largely suppresses spin conduction at low frequencies due to the weakening $J$, thus enhancing DC resistivity across all temperatures. As suggested from results in Fig.~\ref{fig:ACSpin}(c), doping reduces spin resistivity. We find that its trend with temperature is very similar to that at half filling, as can be seen in Fig.~\ref{fig:spinresist}, and close to linear, as predicted due to the lack of spin-charge separation, coupled with the linearity in temperature of the charge resistivity in this region~\cite{q_si_97}.

\begin{figure}[t]
    \includegraphics[width=\linewidth]{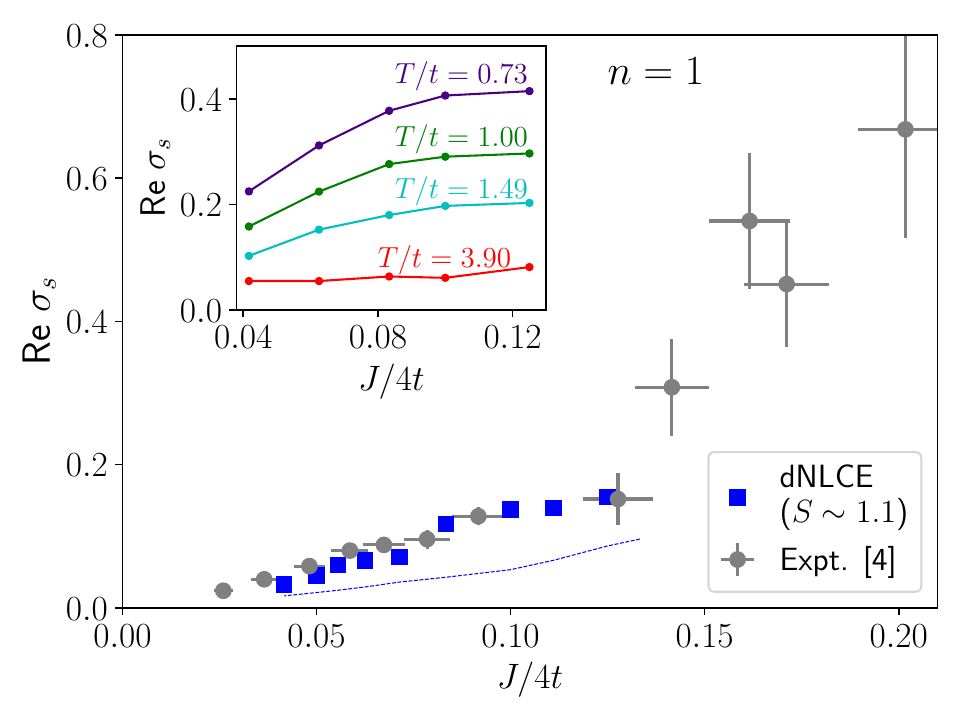}
    \caption{DC spin conductivity of the 2D FHM at half filling as a function of the spin exchange energy, $J/4t=t/U$, at a fixed entropy per particle of $S=1.1$ (main panel), and at several fixed temperatures (inset). Gray points are experimental data from Ref.~\cite{m_nichols_19}. Thin dashed line is the dNLCE result for which the Drude weight is assumed to be zero in the fits to the spin correlation functions.}
    \label{fig:spincond}
\end{figure}

In Fig.~\ref{fig:spincond}, we explicitly look at the dependence of spin conductivity at half filling on the exchange interaction $J$ in the strong-coupling region by plotting an isentropic curve (main panel) and isotherms (inset) in the space of Re $\sigma_s(0)$ and $J/4t=t/U$. In the main panel of Fig.~\ref{fig:spincond}, we find a very good agreement between our results, obtained at a fixed entropy per particle of $S=1.1$ (using exact entropies as functions of temperature calculated in the dNLCE), and data obtained for the same quantity in the optical lattice experiment of Ref.~\cite{m_nichols_19} (gray circles). This constitutes a significant improvement over previous dNLCE estimates for the spin conductivity in the latter study, where the Drude weight was assumed to be negligible on the basis of the temperatures being too high. Here, by explicitly allowing for a nonzero Drude weight via our fits, we find that despite being small and exponentially decreasing with increasing temperature [see Fig.~\ref{fig:ACSpin}(b)], it can lead to large differences in the DC limit of the spin conductivity [see Fig.~\ref{fig:ACCond}(b)], even at high temperatures. When the delta function in the Drude contribution is broadened with the energy scale $J$ to account for the corresponding long-time correlations of the spin current not directly accessible to dNLCE, the Drude weight can account for the large theory-experiment gap previously identified~\cite{m_nichols_19}, at least in the strong-coupling region. 

We also find some evidence that broadening the Drude delta function with a $J$ that is appropriately chosen for the weak-coupling region based on the random phase approximation~\cite{t_paiva_01} can describe the much larger conductivities seen in the experimental data in that region. However, our results at these interaction strengths (not shown) are not as reliable as those in the strong-coupling region. This can be overcome by improving the efficiency of the calculations to access higher orders in the dNLCE expansion in future studies.

\section{Summary and Discussion}
\label{sec:discuss}

While dynamical properties of quantum lattice models are generally challenging to compute, using a combination of out-of-time correlation functions in the real time domain and the NLCE method, which provides access to them at finite temperatures in the thermodynamic limit, we were able to study spin and charge optical conductivity of the 2D FHM at intermediate temperatures and for a range of densities and interaction strengths. Through a combination of fits to capture the long-time behavior of the current correlators and sum rules we were able to estimate the Drude contributions to the DC limit of the conductivities, and demonstrate that using them, one can achieve good agreements with results from optical lattice experiments. We found that the charge resistivity is linear in temperature in the strange metal and the Fermi-liquid regions of the model in a wide temperature range to which we have access. We also found that nonzero Drude weights can account for the difference between experiment and theory results previously observed for spin conductivity at half filling.

Our results in the thermodynamic limit, which are mostly in the strong-coupling region of the FHM, are complementary to those obtained analytically~\cite{d_bergeron_11,mm38-zttx,t_kiely_21} or using the diagramatic Monte Carlo method~\cite{y_eom_25} in the weak- to intermediate-coupling regions.

dNLCE is not limited to conductivities and can potentially be used to compute other dynamical properties and response functions, such as the single-particle spectra~\cite{p_brown_19}, spin and charge dynamical structure factors and susceptibilities~\cite{t_yang_18,r_senaratne_22}, or other response functions~\cite{Prichard2025,l_kendrick_25}. By using the heat current operator of the Hubbard model, one can also use dNLCE to study heat transport of the model and related phenomena~\cite{m_ulaga_22,w_wang_22,w_wang_23,Wang2023,2729-nmyh}. 

dNLCE can be used to obtain conductivities and other dynamical properties for more realistic models of cuprates~\cite{bbyx-gfwl}, e.g., by including nonzero diagonal hopping, or other quantum materials. The calculations can be straightforwardly extended to other geometries and three dimensions too~\cite{triangular_transport,Mendez-Valderrama_21,f_corapi_25}.

Finally, conductivities in the dilute regions are expected to be dominated by the Drude contribution, which can be estimated from time-dependent correlation functions in the dNLCE. The results for conductivities can be compared to those obtained from hydrodynamic equations or other analytical and experimental techniques~\cite{t_kiely_21,PRB_107_155140,f_corapi_25}. 

\acknowledgments
EK is grateful to Martin Zwierlein, Matthew A. Nichols, Lawrence Cheuk, and Thomas Hartke for numerous in-depth discussions about spin conduction of the FHM, including theory calculations, during the completion of the work in Ref.~\cite{m_nichols_19}. We also thank Martin Zwierlein for carefully reviewing the manuscript.
 EK thanks Waseem Bakr for providing experimental data for resistivity in Fig.~\ref{fig:ChargeRes} and for reviewing the manuscript, and Joseph Thywissen for insightful conversations about the charge conductivity of the model, including in the dilute region~\cite{f_corapi_25}. We thank Christoph Karrasch for sharing DMRG results of Ref.~\cite{c_karrasch_14} for the model in 1D for comparisons in Figs.~\ref{fig:dmrgcompare1}-\ref{fig:dmrgcompare2}. We acknowledge support from the U.S. National Science Foundation (NSF) under the Grant No. DMR-1918572. EK's contributions were also supported in part by the NSF Grant No. PHY-2309135 to the Kavli Institute for Theoretical Physics (KITP). Computations were performed on the Spartan high-performance computing facility at San Jos{\'{e} State University supported by the NSF under Grant No. OAC-1626645.

\end{document}